\documentclass[a4paper,onecolumn,11pt,version=2022-06-04]{quantumarticle}
\pdfoutput=1
\usepackage[utf8]{inputenc}
\usepackage[english]{babel}
\usepackage[T1]{fontenc}
\usepackage{amsmath}
\usepackage{amssymb}
\usepackage{hyperref}

\usepackage{tikz}
\usepackage{lipsum}

\newtheorem{theorem}{Theorem}
\newtheorem{definition}{Definition}

\begin{document}

\title{R\'enyi Entropy, Signed Probabilities, and the Qubit}

\author{Adam Brandenburger}
\affiliation{Stern School of Business, Tandon School of Engineering, NYU Shanghai, New York University, New York, NY 10012, U.S.A.}
\email{adam.brandenburger@stern.nyu.edu}
\homepage{adambrandenburger.com}
\author{Pierfrancesco La Mura}
\affiliation{HHL - Leipzig Graduate School of Management, 04109 Leipzig, Germany}
\email{plamura@hhl.de}
\author{Stuart Zoble}
\email{szoble@gmail.com}
\affiliation{Signal Fox, Princeton, NJ 08542, U.S.A.}
\thanks{We are grateful to Samson Abramsky, \v{C}aslav Brukner, Matthew Deutsch, Christopher Fuchs, Leslie Greengard, Michael J.W. Hall, Jerry Keisler, Elliot Lipnowski, Kelvin Onggadinata, David Pine, Rui Soares Barbosa, Blake Stacey, Yuri Tschinkel, and Noson Yanofsky for valuable input.  Financial support from the NYU Stern School of Business, NYU Abu Dhabi, NYU Shanghai, J.P. Valles, and the HHL - Leipzig Graduate School of Management is gratefully acknowledged.}

\maketitle

\begin{abstract}
The states of the qubit, the basic unit of quantum information, are $2\times2$ positive semi-definite Hermitian matrices with trace $1$.  We contribute to the program to axiomatize quantum mechanics by characterizing these states in terms of an entropic uncertainty principle formulated on an eight-point phase space.  We do this by employing R\'enyi entropy (a generalization of Shannon entropy) suitably defined for the signed phase-space probability distributions that arise in representing quantum states.
\end{abstract}

\section{Introduction}
The maximum entropy method was introduced into physics as a way of deriving the Boltzmann distribution of statistical mechanics (Jaynes, 1957).  In this paper, we apply entropy methods to characterize the basic unit of quantum information, namely, the qubit.  Our work fits into the ongoing program to identify principles or axioms yielding quantum mechanics.  This program goes back at least to Birkhoff and von Neumann (1936) and their investigation of quantum mechanics as a non-classical logic.  More recently, Hardy (2001) reconstructed quantum theory from five axioms couched in terms of operations that can be conducted on a physical system.  His work spurred many other axiomatizations based on communication complexity (Van Dam, 2005), information causality (Pawlowski et al., 2009), information capacity (Daki\'c and Brukner, 2011), and purification (Chiribella, et al., 2011), among other principles.

We aim to characterize the simplest quantum system, namely, a two-level system such as the spin of a particle.  Empirically, the experimenter can observe a property such as spin in three arbitrarily chosen mutually orthogonal directions.  In each direction, the outcome is binary (up or down).  An empirical model gives the frequencies of these outcomes when identical copies of the same two-level system are prepared and one of the three measurements is performed on a given copy.  We want to associate an entropy with an empirical model.  This step is not immediate because entropy is a measure of the uncertainty in a single probability distributions, and an empirical model contains three distinct probability distributions (one for each direction).  Our solution is to move to phase space, where an empirical model is represented by a single probability distribution.  The phase space for a two-level system contains eight points, where each point specifies the outcomes (up or down) of each of the three possible measurements.  The possibility of a non-deterministic response to measurement -- as in quantum mechanics -- is allowed for by specifying a probability of each point in phase space.

A phase-space representation of an empirical model can be thought of as a particular (canonical) type of local hidden-variable model (Bell, 1964), where the possible values of the hidden variable are precisely the possible points in phase space.  It follows from Bell's Theorem (Bell, 1964) that there are empirical models which arise in quantum mechanics and which cannot be represented in phase space with ordinary non-negative probabilities.  We follow Wigner (1932), Dirac (1942), and Feynman (1987) in allowing probabilities to be negative, in which case it can be shown that the family of empirical models which can now be represented is precisely the family of all no-signaling theories (Popescu and Rohrlich, 1994; Abramsky and Brandenburger, 2011).  We emphasize that even though the phase-space probabilities are allowed to take negative values, the frequencies of all observable events remain non-negative.

We are now ready to associate entropies with probability distributions on phase space. Within quantum mechanics, the most common entropy measure is the von Neumann entropy (von Neumann, 1932).  This is unsuitable for our purpose because it is defined within the quantum formalism, which we want to derive not assume.  Shannon entropy (Shannon, 1948) is also unsuitable when applied to probabilities in phase space, because the latter can be negative and then Shannon entropy would take complex values.  To avoid these shortcomings, we work with R\'enyi entropy (1961).  This definition of entropy satisfies the basic requirement of extensivity, i.e., it is additive across statistically independent systems.  In fact, it is defined by this property together with some technical axioms (Dar\'oczy, 1963).  R\'enyi entropy is used in various applications in quantum mechanics (Wehner and Winter, 2010; Bialynicki-Birula and Rudnicki, 2011; Coles et al., 2017).

The next step is to state an entropic uncertainty principle as an axiom on phase space.  Note that different from other entropic uncertainty principles in quantum mechanics (Everett, 1957; Hirschman, 1957; Beckner, 1975; Bialynicki-Birula and Mycielski, 1975), our principle is formulated in phase space.  Furthermore, we do not derive the principle from quantum mechanics but introduce it as an axiom.  Our main result is that the set of probability distributions on phase space whose R\'enyi entropy exceeds a certain lower bound is exactly equal to the set of probability distributions that induce the qubit.

We see our axiomatization of the qubit as in line with the program enunciated by Fuchs (2002) to find ``deep \emph{physical} principles'' that yield quantum mechanics.  This said, we do not claim that our axiom is self-evident.  In relativity theory, the principle of light speed invariance is not an intuitive axiom -- the point is that it is physically intelligible.  (This comparison between quantum theory and relativity theory is also made in Onggadinata et al., 2022.)  Our interest in an uncertainty principle is similar.  Ever since the initial formulation by Heisenberg (1927), uncertainty principles have been viewed as one of the notably unintuitive features of quantum mechanics.  But, even though they are mysterious at the everyday macroscopic level, uncertainty principles are physically interpretable, and they are evidently true of microscopic systems.

\section {Preliminaries}

A basis for the space of $2\times2$ Hermitian matrices is given by $\{\boldsymbol{\sigma}_{0},\boldsymbol{\sigma}_{1},\boldsymbol{\sigma}_{2},\boldsymbol{\sigma}_{3}\}$, where $\boldsymbol{\sigma}_{0}=\mathbf{I}$ is the $2\times2$ identity matrix and $\boldsymbol{\sigma}_{1}$, $\boldsymbol{\sigma}_{2}$, $\boldsymbol{\sigma}_{3}$ are the Pauli matrices

\[
\boldsymbol{\sigma}_{1}=\left(\begin{array}{cc}
0 & 1\\
1 & 0
\end{array}\right),\ \boldsymbol{\sigma}_{2}=\left(\begin{array}{cc}
0 & -i\\
i & \phantom{-}0
\end{array}\right),\ \boldsymbol{\sigma}_{3}=\left(\begin{array}{cc}
1 & \phantom{-}0\\
0 & -1
\end{array}\right).
\]

\noindent A $2\times2$ Hermitian matrix $\mathbf{M}$ satisfies $\mbox{Tr}(\mathbf{M})=1$ if and only if

\[
\mathbf{M}=\frac{1}{2}(\mathbf{I}+r_{1}\boldsymbol{\sigma}_{1}+r_{2}\boldsymbol{\sigma}_{2}+r_{3}\boldsymbol{\sigma}_{3})
\]

\noindent for some vector $\boldsymbol{r}=(r_{1},r_{2},r_{3})\in\mathbb{R}^{3}$.

\begin{definition}
A $2\times2$ Hermitian matrix $\mathbf{M}$ with $\mathrm{Tr}(\mathbf{M)}=1$ is called a \emph{potential quantum state}.  If, in addition, $\mathbf{M}$
is positive semi-definite, then $\mathbf{M}$ is a \emph{quantum state}, or a state of the qubit.  We also refer to the corresponding vectors $\boldsymbol{r}$ as potential quantum states and quantum states.
\end{definition}

This is the model of the simplest quantum system, namely a two-level system such as the spin of a particle.  Empirically, the experimenter can observe a property such as spin in three arbitrarily chosen mutually orthogonal directions $x_{1}$, $x_{2}$, and $x_{3}$.  In each direction, the outcome of a measurement will be labeled $+1$ or $-1$.  The expectation of the outcome in direction $i$ is (see, e.g., Sakurai and Napolitano, 2011, p.181)

\[
\mathrm{Tr}(\mathbf{M}\boldsymbol{\sigma}_{i})=r_{i}.
\]

We want to associate an entropy with an empirical model.  This step is not immediate because entropy is a measure of the uncertainty in a single
probability distribution, and an empirical model contains three probability distributions (one for each direction).  Our solution is to move to phase space, where an empirical model is represented by a single probability distribution.  The \emph{phase space} for a two-level system contains eight points,

\[
\{+1,-1\}^{3}=\{\boldsymbol{e}_{n}\ |\ n=1,..,8\},
\]

\noindent where $\boldsymbol{e}_{n}(i)=(-1)^{n_{i}}$ for $(n_{1},n_{2},n_{3})$ the base-$2$ digits of $n-1$.  Each point in phase space specifies the outcome of each of the three possible measurements.  Non-deterministic responses to measurement are incorporated by specifying probabilities over the points in phase space. Let

\[
Q=\{\boldsymbol{q}\in\mathbb{R}^{8}\ |\ \Sigma_{i=1}^{8}q_{i}=1\}
\]

\noindent denote the set of all signed probability distributions on phase space.  That is, we do not require the probabilities to be positive, only that they sum to 1.  We define a map $\phi$ from $Q$ to the set of potential quantum states by
\[
\phi(\boldsymbol{q})=\frac{1}{2}(\mathbf{I}+r_{1}\boldsymbol{\sigma}_{1}+r_{2}\boldsymbol{\sigma}_{2}+r_{3}\boldsymbol{\sigma}_{3}),
\]
\noindent where
\[
r_{i}=\sum_{\{n|\boldsymbol{e}_{n}(i)=+1\}}q_{n}\times(+1)\ +\sum_{\{n|\boldsymbol{e}_{n}(i)=-1\}}q_{n}\times(-1).
\]

\noindent The map $\phi$ gives the correct transformation from phase space to the space of potential quantum states, in the sense of preserving
the empirical probabilities.  This map is linear and it will be helpful to fix some notation surrounding a matrix representation.  Note we have folded the condition that $\boldsymbol{q}$ is a probability distribution in as the last equation in the definition of representation below.

\begin{definition}
Let $\mathbf{A}$ denote the matrix 

{\small{}
\[
\left(\begin{array}{cccccccc}
1 & -1 & \phantom{-}1 & -1 & \phantom{-}1 & -1 & \phantom{-}1 & -1\\
1 & \phantom{-}1 & -1 & -1 & \phantom{-}1 & \phantom{-}1 & -1 & -1\\
1 & \phantom{-}1 & \phantom{-}1 & \phantom{-}1 & -1 & -1 & -1 & -1\\
1 & \phantom{-}1 & \phantom{-}1 & \phantom{-}1 & \phantom{-}1 & \phantom{-}1 & \phantom{-}1 & \phantom{-}1
\end{array}\right).
\]
}{\small\par}

\noindent For $\boldsymbol{r}\in\mathbb{R}^{3}$ define $\boldsymbol{\hat{r}}=(r_{1},r_{2},r_{3},1)\in\mathbb{R}^{4}$.  For $\boldsymbol{q}\in\mathbb{R}^{8}$ and $\boldsymbol{r}\in\mathbb{R}^{3}$ we say $\boldsymbol{q}$ \emph{represents} $\boldsymbol{r}$ if $\mathbf{A}\boldsymbol{q} = \boldsymbol{\hat{r}}$.
\end{definition}

\section {R\'enyi Entropy}
We are going to use phase space to formulate an entropic uncertainty principle as an axiom, and derive the quantum states this way.  In particular, we will allow only those potential quantum states $\boldsymbol{r}$ for which there is a phase-space representation $\boldsymbol{q}$ whose entropy exceeds a lower bound.  The non-classicality of the qubit becomes apparent because there are quantum states for which the only representations with entropy exceeding the bound are signed probability distributions.  The use of negative probabilities on phase space to represent quantum systems goes back to the Wigner quasi-probability probability distribution (Wigner, 1932).  The first task then is to choose a suitable definition of entropy for signed probabilities.

We extend R\'enyi entropy (R\'enyi, 1961) to signed probabilities and impose a smoothness condition that identifies a particular family of entropy functionals.  Fix a finite set $X=\{x_{1},...,x_{n}\}$ together with an ordinary (unsigned) probability distribution $\boldsymbol{q}$ on $X$.  R\'enyi entropy is the family of functionals

\[
H_{\alpha}(\boldsymbol{q})=-\frac{1}{\alpha-1}\log_{2}(\sum_{i=1}^{n}\,q_{i}^{\alpha}),
\]

\noindent where $0<\alpha<\infty$ is a free parameter.  (Shannon entropy is the special case, via L'H\^opital's rule, when $\alpha=1$.)  We can preserve the real-valuedness of entropy under signed probabilities by taking absolute values

\[
H_{\alpha}(\boldsymbol{q})=-\frac{1}{\alpha-1}\log_{2}(\sum_{i=1}^{n}\,|q_{i}|^{\alpha}).
\]

This formula can also be derived axiomatically.  (See Brandenburger and La Mura, 2019, who modify the original axioms for R\'enyi entropy
in R\'enyi, 1961 and Dareczy, 1963.)  We next impose a smoothness condition, requiring that $H_{\alpha}$ be smooth on the space of signed probabilities.  That is, we require $H_{\alpha}(q_1, \ldots, q_{n-1},(1-\sum_{i=1}^{n-1}q_i))$ to be $C^{\infty}$ on $\mathbb{R}^{n-1}$.  Now, if $\alpha$ is not an integer let $k$ be the least integer with $k > \alpha$.  Then

\[
\frac{\partial^{k}H_{\alpha}}{\partial q_{i}^{k}}(q)=\frac{f(q)}{g(q)},
\]

\noindent where $f(q)\neq0$ and $g(q)=0$ for any $q$ with $q_i = 0$.  Thus $\alpha$ must be an integer.  If $\alpha$ is an odd integer then ${\partial^{\alpha}H_{\alpha}}/{\partial q_{i}^{\alpha}}\,(q)$ is undefined for $q$ with $q_i=0$.  Therefore, Rényi entropy takes the following form under our smoothness assumption.

\begin{definition}
\noindent \emph{R\'enyi entropy for signed probability distributions} is the family of functionals

\[
H_{2k}(\boldsymbol{q})=-\frac{1}{2k-1}\log_{2}(\sum_{i=1}^{n}\,q_{i}{}^{2k})=-\frac{2k}{2k-1}\log_{2}(\|\boldsymbol{q}\|_{2k}),
\]

\noindent where $k=1,2,\ldots$ is a free parameter.
\end{definition}

Finally in this section, we give an example of a quantum state such that the only representatives with R\'enyi entropy satisfying the lower bound are signed probabilities.  Consider the quantum state $(r_{1},r_{2},r_{3})=(\frac{1}{\surd3},\frac{1}{\surd3},\frac{1}{\surd3})$.  Set $k=1$.  The (unique) maximum 2-entropy representation is {\small{}

\[
\boldsymbol{q}=\frac{1}{8}(1+\surd3,1+\frac{1}{\surd3},1+\frac{1}{\surd3},1-\frac{1}{\surd3},1+\frac{1}{\surd3},1-\frac{1}{\surd3},1-\frac{1}{\surd3},1-\surd3),
\]

}\noindent with negative final component.  The 2-entropy of $\boldsymbol{q}$ is 2, which is the lower bound we impose below, so we cannot find a representation with all non-negative components with sufficiently high 2-entropy.  In fact any state with $|r_{1}|+|r_{2}|+|r_{3}|>1$ will have this property.

\section{Main Theorem}
We can now state an entropic uncertainty principle as an axiom on phase space.  The axiom is inspired by the use of entropic uncertainty relations in quantum information (Wehner and Winter, 2010; Bialynicki-Birula and Rudnicki, 2011; Coles et al., 2017).

\begin{quote}
\textbf{Uncertainty Principle}: A potential quantum state $\boldsymbol{r}$ satisfies the Uncertainty Principle if for every $k$, there is a phase-space probability distribution $\boldsymbol{q}$ that represents $\boldsymbol{r}$ and satisfies $H_{2k}(\boldsymbol{q})\geq2$.
\end{quote}

\noindent This says that we allow as potential quantum states only those states $\boldsymbol{r}$ containing a minimum amount of uncertainty, as measured by the entropy of a corresponding probability distribution $\boldsymbol{q}$ on phase space.  Note that our Uncertainty Principle is a sequence of conditions, one for each $k$. This is because Rényi entropy itself is not a single functional but a sequence of functionals (indexed by $k$).

\begin{theorem}
The potential quantum states satisfying the Uncertainty Principle are precisely the states of the qubit.
\end{theorem}

\noindent \emph{Proof}. We first show that the potential quantum states satisfying the Uncertainty Principle at $k=1$ are the states of the
qubit. Note that

\[
H_{2}(\boldsymbol{q})\geq2\mbox{ if and only if }\|\boldsymbol{q}\|_{2}^{2}\leq\frac{1}{4}.
\]

\noindent For a general $\boldsymbol{r}$, the representation $\boldsymbol{q}^{*}$ which maximizes $2$-entropy is given by 

\[
\boldsymbol{q}^{*}=\mathbf{A}^{T}(\mathbf{A}\mathbf{A}^{T})^{-1}\boldsymbol{\hat{r}}.
\]

\noindent Using the fact that $\mathbf{A}\mathbf{A}^{T}=8\mathbf{I}$ we have

\[
\|\boldsymbol{q}^{*}\|_{2}^{2}=\boldsymbol{\hat{r}}^{T}(\mathbf{A}\mathbf{A}^{T})^{-1}\boldsymbol{\hat{r}}=\frac{1}{8}\boldsymbol{r}^{T}\boldsymbol{r}+\frac{1}{8}\leq\frac{1}{4}
\]

\noindent if and only if 

\[
r_{1}^{2}+r_{2}^{2}+r_{3}^{2}\leq1,
\]

\noindent and the result follows since the matrix $\frac{1}{2}(\mathbf{I}+r_{1}\boldsymbol{\sigma}_{1}+r_{2}\boldsymbol{\sigma}_{2}+r_{3}\boldsymbol{\sigma}_{3})$
is positive semi-definite if and only if $r_{1}^{2}+r_{2}^{2}+r_{3}^{2}\leq1$.

We now show that if a potential state $\boldsymbol{r}$ satisfies the Uncertainty Principle at $k=1$ then it satisfies the Uncertainty Principle at all $k$.  This is the main mathematical argument.  Fix $k>1$ and let $\boldsymbol{r}\in\mathbb{R}^{3}$ be a state of the qubit.  Choose a $\boldsymbol{q}$ to maximize the $2k$-entropy of a representative of $\boldsymbol{r}$.  We want to show $H_{2k}(\boldsymbol{q})\geq2$ which is equivalent to $\|\boldsymbol{q}\|_{2k}\leq(\frac{1}{2})^{\frac{2k-1}{k}}$.

Observe that $\boldsymbol{q}$ solves the norm minimization problem

\begin{align*}
\min_{\ \boldsymbol{q}\in\mathbb{R}^{8}}\|\boldsymbol{q}\|_{2k}\\
\mathrm{\textrm{subject to}}\  & \mathbf{A}\boldsymbol{q}=\boldsymbol{\hat{r}}.
\end{align*}

\noindent The dual problem is 

\begin{align*}
\max_{\ x\in\mathbb{R}^{4}}\boldsymbol{\hat{r}}^{T}\boldsymbol{x}\\
\ \textrm{subject to}\  & \|\mathbf{A}^{T}\boldsymbol{x}\|_{\frac{2k}{2k-1}}\leq1.
\end{align*}

\noindent (See Boyd and Vandenberghe, 2004, pp.221-222.)  Note that $\|\cdot\|_{\frac{2k}{2k-1}}$ is the dual norm of $\|\cdot\|_{2k}$.  Strong duality holds so the values of the primal and dual problems are equal.  Let $\boldsymbol{y}^{1},\boldsymbol{y}^{k}$ be the maximizers of the dual problems for $2$-entropy and $2k$-entropy respectively.  Let

\[
C_{1}=\{\boldsymbol{x}\in\mathbb{R}^{4}\mid\|\mathbf{A}^{T}\boldsymbol{x}\|_{2}\leq1\}
\]

\noindent and

\[
C_{k}=\{\boldsymbol{x}\in\mathbb{R}^{4}\mid\|\mathbf{A}^{T}\boldsymbol{x}\|_{\frac{2k}{2k-1}}\leq1\}.
\]
\noindent Note that $C_{k}\subseteq C_{1}$ are both convex and, in fact, $C_{1}$ is the ball of radius $\frac{1}{\sqrt{8}}$.  Let 

\[
\boldsymbol{z}^{k}=(\boldsymbol{\hat{r}}^{T}\boldsymbol{y}^{k}/\|\boldsymbol{\hat{r}}\|_{2}^{2})\boldsymbol{\hat{r}}
\]

\noindent be the projection of $\boldsymbol{y}^{k}$ onto $\boldsymbol{\hat{r}}$.  Since $\boldsymbol{\hat{r}}^{T}\boldsymbol{y}^{1}=\frac{\|\boldsymbol{\hat{r}}\|_{2}}{\sqrt{8}}\mbox{cos}\,\theta$, where $\theta$ is the angle between them, we must have $\theta=0$ and so 

\[
\boldsymbol{y}^{1}=(\boldsymbol{\hat{r}}^{T}\boldsymbol{y}^{1}/\|\boldsymbol{\hat{r}}\|_{2}^{2})\boldsymbol{\hat{r}}.
\]

Since the values of the primal and dual problems are equal, these values are positive, so $\frac{\|\boldsymbol{z}^{k}\|_{2}}{\|\boldsymbol{y}^{1}\|_{2}}$ is equal to the ratio of the value of the general $k$ problem to the value of the $k=1$ problem.  By assumption

\[
\boldsymbol{\hat{r}}^{T}\boldsymbol{y}^{1}\leq\frac{1}{2},
\]

\noindent so it is enough to show 

\[
\frac{\|\boldsymbol{z}^{k}\|_{2}}{\|\boldsymbol{y}^{1}\|_{2}}\leq(\frac{1}{2})^{\frac{k-1}{k}}.
\]

We will bound this expression by a function that can be explicitly maximized.  Note that for every nonzero vector $\boldsymbol{w}$ there are unique $\lambda<\nu$ such that 

\[
\|\mathbf{A}^{T}\nu\boldsymbol{w}\|_{2}=1
\]

\noindent and 

\[
\|\mathbf{A}^{T}\lambda\boldsymbol{w}\|_{\frac{2k}{2k-1}}=1.
\]

\noindent This follows immediately from linearity, homogeneity, the fact that $\mathbf{A}$ has full rank, and the fact that $\frac{2k}{2k-1}<2$.  Now let 

\[
f(\boldsymbol{w})=\frac{\|\mathbf{A}^{T}\boldsymbol{w}\|_{2}}{\|\mathbf{A}^{T}\boldsymbol{w}\|_{\frac{2k}{2k-1}}}.
\]
\vspace{0.15in}

\hspace{400bp}

\includegraphics[scale=0.5]{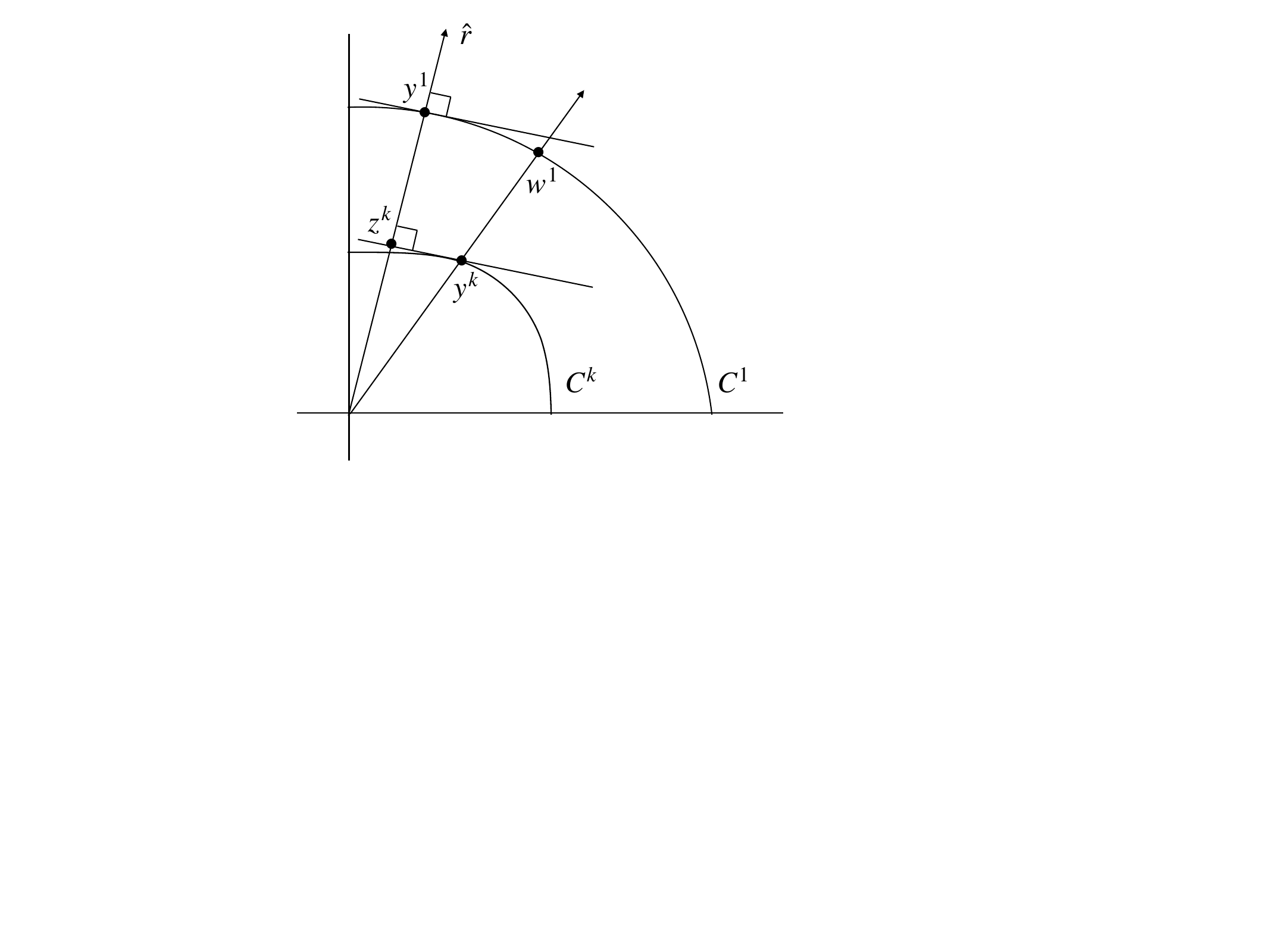}

\vspace{-2.5in}

\hspace{70bp}Figure 1: Comparing the maximizers of the dual problem

\vspace{0.35in}

By the previous observation and the fact that $f(\lambda\boldsymbol{w})=f(\boldsymbol{w})$ for any nonzero scalar $\lambda$, we see that $f(\boldsymbol{w})$ is the ratio of the distance to the boundary of $C_{1}$ along the ray through $\boldsymbol{w}$ to the distance to the boundary of $C_{k}$.  Let $\boldsymbol{w}^{1}=\nu\boldsymbol{y}^{k}$ belong to the boundary of $C_{1}$.  Figure 1 depicts the situation in the plane containing
$\boldsymbol{\hat{r}}$ and $\boldsymbol{y}^{k}$.
\vspace{0.2in}

\noindent \emph{Claim 1}.  $\frac{\|\boldsymbol{z}^{k}\|_{2}}{\|\boldsymbol{y}^{1}\|_{2}}$ is bounded by a value of $f$.

\vspace{0.2in}

\noindent \emph{Proof}. We claim that

\[
\frac{\|\boldsymbol{z}^{k}\|_{2}}{\|\boldsymbol{y}^{1}\|_{2}}\leq\frac{\|\boldsymbol{y}^{k}\|_{2}}{\|\boldsymbol{w}^{1}\|_{2}}.
\]

\noindent Note that 

\[
\boldsymbol{\hat{r}}^{T}\boldsymbol{w}^{1}\leq\boldsymbol{\hat{r}}^{T}\boldsymbol{y}^{1},
\]

\noindent so the length of the projection of $\boldsymbol{w}^{1}$ onto $\boldsymbol{\hat{r}}$ (call this vector $\boldsymbol{v}$) cannot exceed the length of $\boldsymbol{y}^{1}$.  By similar triangles then

\[
\frac{\|\boldsymbol{y}^{k}\|_{2}}{\|\boldsymbol{z}^{k}\|_{2}}=\frac{\|\boldsymbol{w}^{1}\|_{2}}{\|\boldsymbol{v}\|_{2}}\geq\frac{\|\boldsymbol{w}^{1}\|_{2}}{\|\boldsymbol{y}^{1}\|_{2}},
\]

\noindent so $\frac{\|\boldsymbol{z}^{k}\|_{2}}{\|\boldsymbol{y}^{1}\|_{2}}\leq\frac{\|\boldsymbol{y}^{k}\|_{2}}{\|\boldsymbol{w}^{1}\|_{2}}=f(\boldsymbol{w}^{1}).$
{\hfill $\square$}

\vspace{0.2in}

To complete the proof of Theorem 1, it suffices to show that

\[
\max\{f(\boldsymbol{w})\mid\boldsymbol{w}\in\mathbb{R}^{4}\}=(\frac{1}{2})^{\frac{k-1}{k}},
\]

\noindent which we do in the Appendix.

\section{Conclusion}
We have shown that an entropic Uncertainty Principle formulated on an eight-point phase space characterizes the states of the qubit.  We see our result as contributing to the program that aims to reconstruct quantum mechanics from physically interpretable axioms.  Of course, our derivation is only for the simplest, two-level quantum system.  We anticipate that to characterize an $n$-qubit system, methods will be needed that go beyond those in this paper.  In particular, it may be necessary not only to extend our entropic Uncertainty Principle to the $n$-qubit case, but to identify new axioms.  We leave this exploration to future work.

A related derivation of the qubit is Onggadinata et al. (2022).  Similar to our paper, they employ R\'enyi entropy, but fix $\alpha = 2$.  This instance of R\'enyi entropy is often called collision entropy.  Their postulate is that the collision entropy is constant under any dynamics (not necessarily deterministic) on a finite one-dimensional lattice.  From this, they recover the qubit with its full dynamics as defined on the Bloch sphere.

\section*{Appendix}

Let $\boldsymbol{w}\in\mathbb{R}^{4}$ and $\boldsymbol{v}=\boldsymbol{w}\mathbf{A}$.  Let $\boldsymbol{t}\in\mathbb{R}^{8}$ be defined by $t_{i}=v_{i}^{\frac{1}{2k-1}}$.  Note that the critical points of $f$ are the same as the critical points of

\[
\frac{f^{2}(\boldsymbol{w})}{8}=\frac{\|\boldsymbol{v}\|_{2}^{2}}{8\|\boldsymbol{v}\|_{2k/2k-1}^{2}}=\frac{\boldsymbol{w}\mathbf{A}\mathbf{A}^{T}\boldsymbol{w}^{T}}{8\|\boldsymbol{v}\|_{2k/2k-1}^{2}}=\frac{\|\boldsymbol{w}\|_{2}^{2}}{\|\boldsymbol{v}\|_{2k/2k-1}^{2}},
\]

\noindent which are the solutions of the system of first-order conditions

\[
w_{i}=h(\boldsymbol{w})r_{i}\boldsymbol{t}^{T}\ \ i=1,2,3,4,
\]

\noindent where 

\[
h(\boldsymbol{w})=\frac{\|\boldsymbol{w}\|_{2}^{2}}{\|\boldsymbol{v}\|_{2k/2k-1}^{2k/2k-1}}>0
\]

\noindent and $r_{i}$ is the $i$th row of the matrix $\mathbf{A}$.  It is helpful to write out the system with $\gamma$ denoting $\frac{1}{2k-1}$
for readability:
\small

\[
w_{1}=h(\boldsymbol{w})[(w_{1}+w_{2}+w_{3}+w_{4})^{\gamma}-(-w_{1}+w_{2}+w_{3}+w_{4})^{\gamma}+(w_{1}-w_{2}+w_{3}+w_{4})^{\gamma}-(-w_{1}-w_{2}+w_{3}+w_{4})^{\gamma}+
\]
\[
(w_{1}+w_{2}-w_{3}+w_{4})^{\gamma}-(-w_{1}+w_{2}-w_{3}+w_{4})^{\gamma}+(w_{1}-w_{2}-w_{3}+w_{4})^{\gamma}-(-w_{1}-w_{2}-w_{3}+w_{4})^{\gamma}],
\]
\vspace{0.05in}
\[
w_{2}=h(\boldsymbol{w})[(w_{1}+w_{2}+w_{3}+w_{4})^{\gamma}+(-w_{1}+w_{2}+w_{3}+w_{4})^{\gamma}-(w_{1}-w_{2}+w_{3}+w_{4})^{\gamma}-(-w_{1}-w_{2}+w_{3}+w_{4})^{\gamma}+
\]
\[
(w_{1}+w_{2}-w_{3}+w_{4})^{\gamma}+(-w_{1}+w_{2}-w_{3}+w_{4})^{\gamma}-(w_{1}-w_{2}-w_{3}+w_{4})^{\gamma}-(-w_{1}-w_{2}-w_{3}+w_{4})^{\gamma}],
\]
\vspace{0.05in}
\[
w_{3}=h(\boldsymbol{w})[(w_{1}+w_{2}+w_{3}+w_{4})^{\gamma}+(-w_{1}+w_{2}+w_{3}+w_{4})^{\gamma}+(w_{1}-w_{2}+w_{3}+w_{4})^{\gamma}+(-w_{1}-w_{2}+w_{3}+w_{4})^{\gamma}-
\]
\[
(w_{1}+w_{2}-w_{3}+w_{4})^{\gamma}-(-w_{1}+w_{2}-w_{3}+w_{4})^{\gamma}-(w_{1}-w_{2}-w_{3}+w_{4})^{\gamma}-(-w_{1}-w_{2}-w_{3}+w_{4})^{\gamma}],
\]
\vspace{0.05in}
\[
w_{4}=h(\boldsymbol{w})[(w_{1}+w_{2}+w_{3}+w_{4})^{\gamma}+(-w_{1}+w_{2}+w_{3}+w_{4})^{\gamma}+(w_{1}-w_{2}+w_{3}+w_{4})^{\gamma}+(-w_{1}-w_{2}+w_{3}+w_{4})^{\gamma}+
\]
\[
(w_{1}+w_{2}-w_{3}+w_{4})^{\gamma}+(-w_{1}+w_{2}-w_{3}+w_{4})^{\gamma}+(w_{1}-w_{2}-w_{3}+w_{4})^{\gamma}+(-w_{1}-w_{2}-w_{3}+w_{4})^{\gamma}].
\]
\normalsize
\vspace{0.05in}

\noindent \emph{Claim 2}. The system $\boldsymbol{w}=h(\boldsymbol{w})\mathbf{A}\boldsymbol{t}^{T}$ has the following properties:

\begin{enumerate}
\item If $\boldsymbol{w}$ is a solution then so is $\lambda\boldsymbol{w}$ for any $\lambda\neq0$.
\item If $\boldsymbol{w}$ is a solution then $\boldsymbol{v}$ is a solution, where $\boldsymbol{v}$ is obtained from $\boldsymbol{w}$ by permuting coordinates.
\end{enumerate}

\noindent \emph{Proof}.  For (1) we have $h(\lambda\boldsymbol{w})\boldsymbol{t}^{T}(\lambda\boldsymbol{w})=\frac{\lambda^{2}\lambda^{1/2k-1}}{\lambda^{2k/2k-1}}h(\boldsymbol{w})\boldsymbol{t}^{T}=\lambda\boldsymbol{w}$.  For (2) we have 

\[
w_{1}=h(w_{4},w_{2},w_{3},w_{1})r_{4}\boldsymbol{t}^{T}(w_{4},w_{2},w_{3},w_{1})
\]
\noindent and

\[
w_{4}=h(w_{4},w_{2},w_{3},w_{1})r_{1}\boldsymbol{t}^{T}(w_{4},w_{2},w_{3},w_{1}),
\]

\noindent and similarly for $w_{2},w_{3}$.
{\hfill $\square$}
\vspace{0.05in}

\noindent \emph{Claim 3}.  Assume $w_{4}\neq0$. Let $i,j<4$.  Then 

\[
|w_{i}|=|w_{j}|\mbox{ or }w_{i}w_{j}=0.
\]

\noindent{Proof}.  We may assume $w_{4}>0$.  For $N$ sufficiently large we have $\|\boldsymbol{w}-\boldsymbol{a}\|_{2}<\|\boldsymbol{a}\|_{2}$ where $\boldsymbol{a}=(0,0,0,N)$.  Thus the Taylor series expansion of $r_{i}\boldsymbol{t}^{T}$ at the point $\boldsymbol{a}$ converges at $\boldsymbol{w}$.  We have

\[
w_{1}=8h(\boldsymbol{w})\sum\limits _{\substack{\alpha_{1}\in O\\
\alpha_{2},\alpha_{3}\in E\\
\alpha_{4}\in\mathbb{N}
}
}\frac{(\boldsymbol{w}-\boldsymbol{a})^{\boldsymbol{\alpha}}}{\boldsymbol{\alpha}!}C(\sum_{i=1}^{4}\alpha_{i}),
\]
\vspace{0.05in}
\[
w_{2}=8h(\boldsymbol{w})\sum\limits _{\substack{\alpha_{2}\in O\\
\alpha_{1},\alpha_{3}\in E\\
\alpha_{4}\in\mathbb{N}
}
}\frac{(\boldsymbol{w}-\boldsymbol{a})^{\boldsymbol{\alpha}}}{\boldsymbol{\alpha}!}C(\sum_{i=1}^{4}\alpha_{i}),
\]
\vspace{0.05in}
\[
w_{3}=8h(\boldsymbol{w})\sum\limits _{\substack{\alpha_{3}\in O\\
\alpha_{1},\alpha_{2}\in E\\
\alpha_{4}\in\mathbb{N}
}
}\frac{(\boldsymbol{w}-\boldsymbol{a})^{\boldsymbol{\alpha}}}{\boldsymbol{\alpha}!}C(\sum_{i=1}^{4}\alpha_{i}),
\]
\vspace{0.05in}
\[
w_{4}=8h(\boldsymbol{w})\sum\limits _{\substack{\alpha_{1},\alpha_{2},\alpha_{3}\in E\\
\alpha_{4}\in\mathbb{N}
}
}\frac{(\boldsymbol{w}-\boldsymbol{a})^{\boldsymbol{\alpha}}}{\boldsymbol{\alpha}!}C(\sum_{i=1}^{4}\alpha_{i}),
\]

\noindent where $\boldsymbol{\alpha}\in\mathbb{N}^{4}$ is a multi-index, $\mathbb{N}=E\cup O=\{0,2,4,...\}\cup\{1,3,5,...\},$

\[
\boldsymbol{\alpha}!=\alpha_{1}!\alpha_{2}!\alpha_{3}!\alpha_{4}!,
\]
\[
(\boldsymbol{w}-\boldsymbol{a})^{\boldsymbol{\alpha}}=w_{1}^{\alpha_{1}}w_{2}^{\alpha_{2}}w_{3}^{\alpha_{3}}(w_{4}-N)^{\alpha_{4}},
\]

\noindent and $C$ is defined by

\[
C(0)=1,\,C(1)=\frac{1}{(2k-1)N^{\frac{2k}{2k-1}}},\,{\rm and}
\]
\vspace{0.05in}
\[
C(n)=\frac{(-1)^{n-1}\prod_{j=1}^{n-1}(j(2k-1)-1)}{(2k-1)^{n}N^{\frac{n(2k-1)-1}{2k-1}}}\ \mbox{ for }n>1.
\]

Note that $C(\sum_{i=1}^{4}\alpha_{i})>0$ if and only if $\sum_{i=1}^{4}\alpha_{i}\in O$.  Assume $w_{1},w_{2}\neq0$.  We have

\[
w_{1}\sum\limits _{\substack{\alpha_{2}\in O\\
\alpha_{1},\alpha_{3}\in E\\
\alpha_{4}\in\mathbb{N}
}
}\frac{(\boldsymbol{w}-\boldsymbol{a})^{\boldsymbol{\alpha}}}{\boldsymbol{\alpha}!}C(\sum_{i=1}^{4}\alpha_{i})=w_{2}\sum\limits _{\substack{\alpha_{1}\in O\\
\alpha_{2},\alpha_{3}\in E\\
\alpha_{4}\in\mathbb{N}
}
}\frac{(\boldsymbol{w}-\boldsymbol{a})^{\boldsymbol{\alpha}}}{\boldsymbol{\alpha}!}C(\sum_{i=1}^{4}\alpha_{i}),
\]

\noindent equivalently 
\[
\sum\limits _{\substack{\alpha_{2}\in O\\
\alpha_{1},\alpha_{3}\in E\\
\alpha_{4}\in\mathbb{N}
}
}\frac{(\boldsymbol{w}-\boldsymbol{a})^{\boldsymbol{\alpha}+(1,0,0,0)}}{\boldsymbol{\alpha}!}C(\sum_{i=1}^{4}\alpha_{i})=\sum\limits _{\substack{\alpha_{1}\in O\\
\alpha_{2},\alpha_{3}\in E\\
\alpha_{4}\in\mathbb{N}
}
}\frac{(\boldsymbol{w}-\boldsymbol{a})^{\boldsymbol{\alpha}+(0,1,0,0)}}{\boldsymbol{\alpha}!}C(\sum_{i=1}^{4}\alpha_{i}).
\]

Re-indexing we have 

\[
\sum\limits _{\substack{\alpha_{1},\alpha_{2}\in O\\
\alpha_{3}\in E\\
\alpha_{4}\in\mathbb{N}
}
}\alpha_{1}\frac{(\boldsymbol{w}-\boldsymbol{a})^{\boldsymbol{\alpha}}}{\boldsymbol{\alpha}!}C(\sum_{i=1}^{4}\alpha_{i}-1)=\sum\limits _{\substack{\alpha_{1},\alpha_{2}\in O\\
\alpha_{3}\in E\\
\alpha_{4}\in\mathbb{N}
}
}\alpha_{2}\frac{(\boldsymbol{w}-\boldsymbol{a})^{\boldsymbol{\alpha}}}{\boldsymbol{\alpha}!}C(\sum_{i=1}^{4}\alpha_{i}-1).
\]

Collecting terms we have

\[
\sum\limits _{\substack{\alpha_{1}<\alpha_{2}\\
\alpha_{1},\alpha_{2}\in O\\
\alpha_{3}\in E\\
\alpha_{4}\in\mathbb{N}
}
}(\alpha_{1}-\alpha_{2})(w_{1}^{\alpha_{1}}w_{2}^{\alpha_{2}}-w_{1}^{\alpha_{2}}w_{2}^{\alpha_{1}})w_{3}^{\alpha_{3}}(w_{4}-N)^{\alpha_{4}}\frac{C(\sum_{i=1}^{4}\alpha_{i}-1)}{\boldsymbol{\alpha}!}=0,
\]

\noindent equivalently

\[
\sum\limits _{\substack{\alpha_{1}<\alpha_{2}\\
\alpha_{1},\alpha_{2}\in O\\
\alpha_{3}\in E\\
\alpha_{4}\in\mathbb{N}
}
}(\alpha_{1}-\alpha_{2})w_{1}^{\alpha_{1}}w_{2}^{\alpha_{2}}(1-(\frac{w_{1}}{w_{2}})^{\alpha_{2}-\alpha_{1}})w_{3}^{\alpha_{3}}(w_{4}-N)^{\alpha_{4}}\frac{C(\sum_{i=1}^{4}\alpha_{i}-1)}{\boldsymbol{\alpha}!}=0.
\]

The key point is that 

\[
(\alpha_{1}-\alpha_{2})w_{1}^{\alpha_{1}}w_{2}^{\alpha_{2}}w_{3}^{\alpha_{3}}(w_{4}-N)^{\alpha_{4}}\frac{C(\sum_{i=1}^{4}\alpha_{i}-1)}{\boldsymbol{\alpha}!}
\]

\noindent always has the same sign as $w_{1}w_{2}$.  Thus since $\alpha_{2}-\alpha_{1}\in E$ we conclude that $|w_{1}|=|w_{2}|$.
{\hfill $\square$}
\vspace{0.07in}

\noindent \emph {Claim 4}.  If $w_{i},w_{j}\neq0$ then $|w_{i}|=|w_{j}|$ .
\vspace{0.07in}

\noindent{Proof}.  By Claim 2 and Claim 3 we may assume that $w_{1},w_{4}\neq0$ and $w_{2},w_{3}=0$.  We may further assume that $w_{4}=1$.  Thus the equation for $w_{1}$ becomes 

\[
w_{1}=\frac{(w_{1}+1)^{1/2k-1}-(-w_{1}+1)^{1/2k-1}}{(w_{1}+1)^{1/2k-1}+(-w_{1}+1)^{1/2k-1}}
\]

\noindent so 

\[
(w_{1}+1)(-w_{1}+1)^{1/2k-1}=(-w_{1}+1)(w_{1}+1)^{1/2k-1},
\]

\noindent from which we conclude that $w_{1}\in\{-1,1\}$ as desired.
{\hfill $\square$}
\vspace{0.05in}

We have thus shown that for every $i,j\leq4$ either $|w_{i}|=|w_{j}|$ or $w_{i}w_{j}=0$ so we need only consider critical points with 

\[
w_{i}\in\{0,1,-1\}
\]

\noindent for each $i=1,...,4$.  It is easy to check that the maximum of the original function $f$ occurs when exactly two of the weights are $0$ and this maximum value is $(\frac{1}{2})^{\frac{k-1}{k}}$, completing the proof of Theorem 1.

\section*{References}

\noindent Abramsky, S., and A. Brandenburger, ``The Sheaf-Theoretic Structure of Non-Locality and Contextuality,'' \emph{New Journal of Physics}, 13, 2011, 113036.\\

\noindent Beckner, W., ``Inequalities in Fourier Analysis,'' \emph{Annals of Mathematics}, 102, 1975, 159-182.\\

\noindent Bell, J., ``On the Einstein-Podolsky-Rosen Paradox,'' \emph{Physics}, 1, 1964, 195-200.\\

\noindent Bialynicki-Birula, I., and J. Mycielski, ``Uncertainty Relations for Information Entropy in Wave Mechanics,'' \emph{Communications in Mathematical Physics}, 44, 1975, 129-132.\\

\noindent Bialynicki-Birula, I., and \L. Rudnicki, ``Entropic Uncertainty Relations in Quantum Physics,'' in K. Sen (ed.), \emph{Statistical Complexity: Applications in Electronic Structure}, Springer, 2011, 1-34.\\

\noindent Birkhoff, G., and J. von Neumann, ``The Logic of Quantum Mechanics,'' \emph{Annals of Mathematics}, 37, 1936, 823-843.\\

\noindent Boyd, S., and L. Vandenberghe, \emph{Convex Optimization}, Cambridge University Press, 2004.\\

\noindent Brandenburger, A., and P. La Mura, ``Axioms for Rényi Entropy with Signed Measures,'' 2019, at http://www.adambrandenburger.com.\\

\noindent Chiribella, G., G. D’Ariano, and P. Perinotti, ``Informational Derivation of Quantum Theory,'' \emph{Physical Review A}, 84, 2011, 012311.\\

\noindent Coles, P., M. Beta, M. Tomamichel, and S. Wehner, ``Entropic Uncertainty Relations and Their Applications,'' \emph{Reviews of Modern Physics}, 89, 2017, 015002.\\

\noindent Daki\'c, B., and C. Brukner, ``Quantum Theory and Beyond: Is Entanglement Special?'' in Halvorson, H. (ed.), \emph{Deep Beauty: Understanding the Quantum World through Mathematical Innovation}, Cambridge University Press, 2011, 365-392.\\

\noindent Dar\'{o}czy, Z., ``{\"U}ber die gemeinsame Charakterisierung der zu den nicht vollständigen Verteilungen gehörigen Entropien von Shannon und von Rényi,'' \emph{Zeitschrift für Wahrscheinlichkeitstheorie und verwandte Gebiete}, 1, 1963, 381-388.\\

\noindent Everett, H., \emph{On the Foundations of Quantum Mechanics}, doctoral dissertation, Princeton University, 1957.\\

\noindent Fuchs, C., “Quantum Mechanics as Quantum Information (and Only a Little More),” in A. Khrennikov, (ed.), Quantum Theory: Reconsideration of Foundations, V\"{a}xj\"{o} University Press, 2002, 463-543.\\

\noindent Heisenberg, W., ``\"{U}ber den anschulichen Inhalt der quantentheoretischen Kinematik und Mechanik,'' \textit{Zeitschrift f\"{u}r Physik}, 43, 1927, 172-198.\\

\noindent Hirschman, I., ``A Note on Entropy,'' \emph{American Journal of Mathematics}, 79, 1957, 152-156.\\

\noindent Onggadinata, K., P. Kurzynski, and D. Kaszlikowski, ``Qubit from the Classical Collision Entropy,'' 2022, at https://arxiv.org/abs/2205.00773.\\

\noindent Pawlowski M., T. Paterek, D. Kaszlikowski, V. Scarani, A. Winter, and M. Zukowski, ``Information Causality as a Physical Principle,'' \emph{Nature}, 461, 2009, 1101-1104.\\

\noindent Popescu, S., and D. Rohrlich, ``Quantum Nonlocality as an Axiom,'' \emph{Foundations of Physics}, 24, 1994, 379-385.\\

\noindent Rényi, A., ``On Measures of Information and Entropy,'' in J. Neymann (ed.), \emph{Proceedings of the 4th Berkeley Symposium on Mathematical Statistics and Probability}, University of California Press, 1961, 547-561.\\

\noindent Sakurai, J.J., and J. Napolitano, \emph{Modern Quantum Mechanics}, Addison-Wesley, 2nd edition, 2011.\\

\noindent Shannon, C., ``A Mathematical Theory of Communication,'' \emph{Bell System Technical Journal}, 27, 1948, 379-423 and 623-656.\\

\noindent Van Dam, W., ``Implausible Consequences of Superstrong Nonlocality,'' 2005, at http://arxiv.\break org/abs/quant-ph/0501159.\\
\newpage
\noindent Von Neumann, J., \emph{Mathematische Grundlagen der Quantenmechanik}, Springer, Berlin, 1932.\\

\noindent Wehner, S., and A. Winter, ``Entropic Uncertainty Relations -- A Survey,'' \emph{New Journal of Physics}, 12, 2010, 025009.\\

\noindent Wigner, E., ``On the Quantum Correction For Thermodynamic Equilibrium,'' \emph{Physical Review}, 40, 1932, 749-759.\\

\end{document}